\documentstyle[12pt]{article}
\input epsf.tex

\begin{document}

\baselineskip=17pt
\pagestyle{plain}
\setcounter{page}{1}

\begin{titlepage}

\begin{flushright}
PUPT-1760 \\
hep-th/9803097
\end{flushright}
\vspace{20 mm}

\begin{center}
{\huge Worldsheet Dynamics of String Junctions}

\vspace{5mm}

\end{center}

\vspace{15 mm}

\begin{center}
{\large Curtis G.\ Callan, Jr.\footnote{callan@viper.princeton.edu}
and L\'arus Thorlacius\footnote{larus@feynman.princeton.edu}}

\vspace{3mm}

Joseph Henry Laboratories\\
Princeton University\\
Princeton, New Jersey 08544
\end{center}

\vspace{1.5cm}

\begin{center}
{\large Abstract}
\end{center}

\noindent
We analyze scattering of string modes at string junctions of type IIB string 
theory. In the infrared limit, certain orthogonal linear combinations of the 
fields on the different strings satisfy either Dirichlet or Neumann boundary 
conditions.  We confirm that the worldsheet theory of a general string 
network has eight conserved supercharges in agreement with target space BPS 
considerations. As an application, we obtain the band spectrum of some 
simple string lattices.

\vspace{2cm}
\begin{flushleft}
March 1998
\end{flushleft}
\end{titlepage}
\newpage
\renewcommand{\baselinestretch}{1.1}  %looks better

%%%%%%%%%%%%%%%%%%%%%%%%%%%%%%%%%%%%%%%%%%%%%%
%% include the next line for double spacing %%
%%%%%%%%%%%%%%%%%%%%%%%%%%%%%%%%%%%%%%%%%%%%%%
%\renewcommand{\baselinestretch}{2}

\newcommand{\half}{{1\over 2}}
\newcommand{\third}{{1\over 3}}
\newcommand{\be}{\begin{equation}}
\newcommand{\ee}{\end{equation}}
\newcommand{\bea}{\begin{eqnarray}}
\newcommand{\eea}{\end{eqnarray}}

\section{Introduction}
As several authors have recently pointed out, triple junctions of
type IIB $(p,q)$ strings \cite{asy,schwarz} preserve supersymmetry and can 
be used to build supersymmetric networks of junctions \cite{sensusy,mukhi}.
Other applications include gauge symmetry enhancement to exceptional 
groups in type IIB string theory \cite{gabzwi} and description of $\nu=1/4$ 
BPS states in $N=4$ supersymmetric Yang-Mills theories \cite{bergman}. 
Natural questions arise concerning the behavior of excitations of 
such networks: What is the S-matrix describing the scattering from a 
single junction? Are there BPS excited states of networks laid out on 
compact tori and, if so, what can we say about their entropy? 

In this paper, we will explore what can be learned from an extremely 
simple-minded view of the problem. The individual $(p,q)$ strings all support
the standard massless multiplet of 8 bosons and 8 fermions (both left-moving
and right-moving). The triple junction is a common boundary to the three
string worldsheets at which the left-moving excitations on any one string
can scatter into right-moving excitations on all of them. More complicated 
things can happen at high energies, but at energies well below the string 
scale the scattering should be linear (an ``in'' quantum of given energy on
one string should scatter into a linear combination of ``out'' quanta on
all the strings entering the junction). 

As we will show, rather simple physical arguments suffice to determine the 
linearized junction S-matrix.  The reflection and transmission amplitudes
for modes on the individual strings that extend from the junction depend
on the tensions and relative orientation of the strings.  It turns out,
however, that the S-matrix can always be diagonalized to give standard
Dirichlet or Neumann boundary conditions for particular combinations of
modes on the different strings.  We carry out the analysis both for 
fluctuations that are transverse to the plane of the string junction and
for `in-plane' modes.  We then generalize the discussion to include 
higher order junctions where $n>3$ co-planar strings meet.

The worldsheet theory of a general BPS network of $(p,q)$ strings is 
expected to have eight of the thirty-two supergenerators of type~IIB
string theory unbroken.  We verify that the S-matrices at the various
string junctions in the network indeed preserve precisely the right
number of supercharges.  

We close with a discussion of the excitation spectrum of a periodic
string lattice.  Sen has proposed that such lattices, viewed as string
networks laid out on tori, can be used as novel building blocks for
string compactification \cite{sensusy}.  Carrying out this idea in 
practice will require some understanding of the basic dynamical 
properties of triple string junctions. Initial steps in that direction
have been taken by Rey and Yee \cite{soojong}. Our line of approach differs 
from theirs and gives, we believe, a more detailed understanding of the 
dynamics. 

\section{Transverse Mode Scattering at a Junction}

Let us consider a junction where three strings meet, as shown
in Figure~1.  We'll 
generalize to higher order string junctions in Section~4.  We take the
strings to lie in the complex plane with the equilibrium position of
the junction at $z=0$ and the strings making angles $\theta_i$ with
the positive real axis.  The three strings have NS-NS and R-R charges 
$p_i$ and $q_i$ respectively.  
\vskip .7cm
\vbox{
{\centerline{\epsfxsize=2.4in \epsfbox{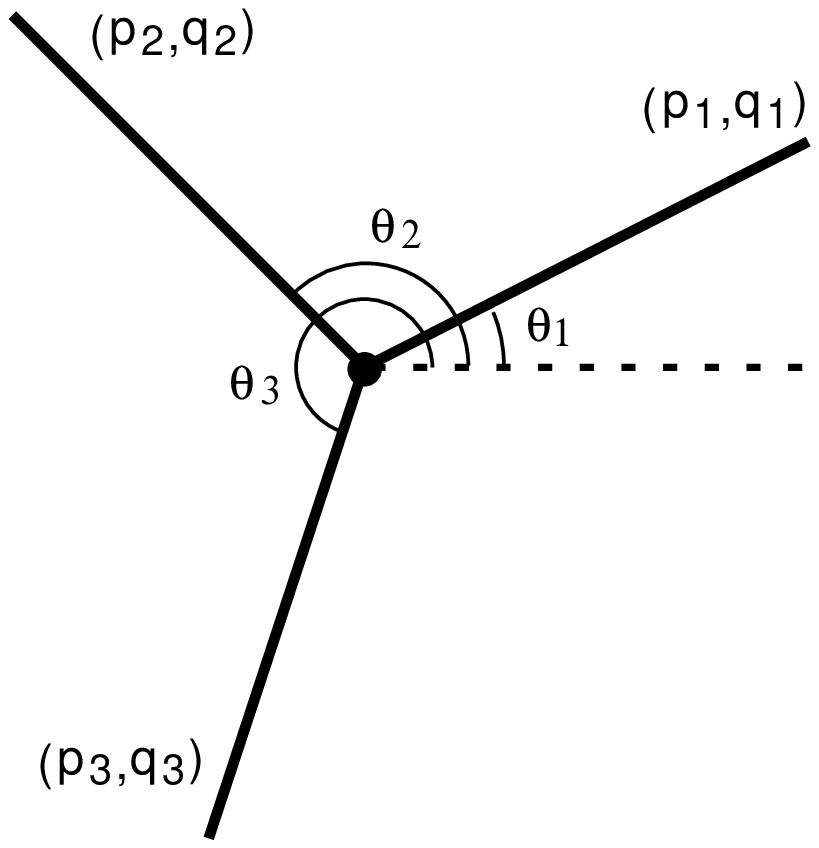}}}
{\centerline{ FIGURE 1:
A three-string junction.}}
}
\vskip .5cm
The configuration is stable and preserves
$\nu=1/4$ of the type IIb supersymmetry if the charges are conserved
at the junction and each string is aligned with its charge vector
$p_i+q_i\tau$, where $\tau$ is the usual axion-dilaton modulus of
type IIb theory \cite{sensusy,mukhi}.  Charge conservation and force 
balance are then expressed as
\be
\label{conserve}
0=\sum_{i=1}^3 q_i = \sum_{i=1}^3 p_i = \sum_{i=1}^3 e^{i\theta_i}t_i ,
\ee
where $\theta_i$ is the argument of $p_i+q_i \tau$:
\be
\label{pplusq}
p_i+q_i\tau = \vert p_i+q_i\tau \vert e^{i\theta_i},
\ee
and $t_i$ is the scalar string tension,
\be
t_i = {1\over \sqrt{\vert{\rm Im}\tau\vert}} \vert p_i + q_i \tau \vert .
\label{tensioni}
\ee

We first consider fluctuations $\phi_i$ that are transverse to the plane of 
the three-string junction.  Here the subscript $i=1,2,3$ denotes the string 
on which the fluctuation is found.  There are seven independent transverse
fluctuations on each string. The scattering problem is diagonal in this 
flavor space and we have suppressed the corresponding indices on $\phi$. 
Since we are dealing with real massless fields, the general expression for 
a mode of frequency $\omega$ can be written
\be
\phi_i(x_i,t) = Re\{(A_i e^{i\omega x_i} + B_i e^{-i\omega x_i})
	e^{-i\omega t}\} ,
\label{modes}
\ee
where $A_i,B_i$ are complex mode amplitudes and $x_i>0$ is the distance 
from the junction measured along the given string. 
The physical matching conditions at $x_i=0$ are continuity, expressed as
\begin{equation}
\phi_1(0) = \phi_2(0) = \phi_3(0) ,
\label{match}
\end{equation}
and `vertical' ({\it i.e.} transverse to plane of junction) tension balance,
\begin{equation}
\sum_{i=1}^3  t_i \, \phi_i '(0) =0~.
\label{tbal}
\end{equation}
In these equations, we suppress the time argument: they are supposed to
hold at all times. As is to be expected, the matching conditions treat 
the three strings in a completely symmetric fashion. 

A string junction has an M-theory description in terms of a 
three-pronged wrapped membrane in ${\bf R}^9\times {\bf T}^2$
\cite{mkrogh,matsuo}.
Our matching conditions can easily be derived in that context.  The
continuity condition follows immediately from the fact that the 
junction is described by a single membrane.  The tension balance 
condition requires more work, but it follows from a variational
calculation, involving the Nambu-Goto action for the membrane,
applied to a pants-like section of the membrane that includes the
junction and connects to each of the extended strings.  We will 
not go into the details here but simply consider the condition
(\ref{tbal}) to be physically well-motivated and proceed to work
out the resulting scattering problem.

It is easy to solve the boundary conditions to find the $3\times 3$ 
scattering matrix relating the ``in" and ``out" modes: 
$\vec A=S\cdot \vec B$, where 
\be
{ S}=-{\bf 1} +
{2\over \sum_1^3 t_i}
\left(\begin{array}{ccc}
t_1& t_2& t_3\cr
t_1& t_2& t_3\cr
t_1& t_2& t_3
\end{array}\right) \,.
\label{smat}
\ee
The matrix $S$ is real valued and therefore it acts independently
on the real and imaginary parts of the complex mode amplitudes.
It has eigenvalues $\pm 1$, with one $+1$ eigenvector
and two $-1$ eigenvectors:
$S\cdot(1,1,1)= (1,1,1)$ and $S\cdot\vec x=-\vec x$ if 
$\vec t\cdot\vec x=0$,
where $\vec t=(t_1,t_2,t_3)$ (there are two such vectors).
%This is the only way the S-matrix $\vec A^*=S\cdot \vec A$ can be consistent:
%the real part of $\vec A$ must be proportional to the $+1$ eigenvector
%and the imaginary part must be in the space spanned by the $-1$ eigenvectors.
%We can check that energy is conserved in this scattering process:

The action which describes the dynamics of the fields $\phi_i$ is 
\be
L = \sum_i t_i \int_0^\infty dx_i(\partial\phi_i)^2~.
\label{action}
\ee
If we supplement it with the constraint that the three fields be equal
at the origin, it contains all the boundary conditions discussed above. 
For a quadratic action like this, we can eliminate the tensions from the 
action and the energy by rewriting them in terms of the rescaled fields
$\hat \phi_i=\sqrt{t_i}\phi_i$. On the other hand, the tensions then appear 
in the boundary conditions in a slightly more complicated way:
\be
{\hat\phi_1(0)\over \sqrt{t_1}}={\hat\phi_2(0)\over \sqrt{t_2}}
={\hat\phi_3(0)\over \sqrt{t_3}} ~, 
\qquad \sum_{i=1}^3\sqrt{t_i}\,\hat\phi_i^\prime(0)=0~.
\label{hatbc}
\ee
The S-matrix for the rescaled fields is very simply related to the old one:
$\hat S =\sqrt{t} S \sqrt{t}^{-1}$ where $t$ is a diagonal matrix whose 
entries are the scalar tensions. Recalling (\ref{smat}), we find 
\be
\hat S = -{\bf 1} + 2 \vec y\otimes \vec y ~,
\label{shat}
\ee
where $\vec y$ is the unit three-vector
\be
\vec y = {1\over\sqrt{\sum{t_i}}}(\sqrt{t_1},\sqrt{t_2},\sqrt{t_3}) ~.
\label{yvect}
\ee
The rescaled S-matrix is symmetric and squares to the identity. 
It has eigenvalues $\pm1$ and its eigenvectors are orthogonal. 
It follows immediately that energy is conserved in the scattering 
process. In terms of rescaled fields, the energy carried by the incoming 
modes is $E_{in}=\sum_{i=1}^3 \omega^2 \vert\hat{\vec B}\vert^2$, 
while the energy carried by the outgoing modes is the same thing with 
$\hat{\vec B}\to \hat{\vec A}$. 

The eigenvalues $+1$ and $-1$ correspond to Neumann and Dirichlet boundary 
conditions respectively.  The $+1$ eigenvector is $\vec y$ and any vector 
normal to $\vec y$ is a $-1$ eigenvector.  Thus, we can ``trivialize'' the 
scattering problem by taking the appropriate linear combinations of the fields 
on the three different strings! Note that, according
to (\ref{modes}), only the mode with +1 eigenvalue can correspond to a zero 
mode (a mode whose amplitude does not vanish in the $\omega\to 0$ limit). 
We easily see from (\ref{smat}) that this mode has $A_1=A_2=A_3$, which
is to say that it amounts to a uniform translation of the junction
transverse to its plane. This is precisely the zero mode we would expect. 

To further check that these results correspond to expectations, 
consider a $(1,0)$ string ({\it i.e.} a fundamental string) attached to 
$(0,q)$ and $(-1,-q)$ strings at weak string coupling.  
In this case $t_1 << t_2 \sim t_3$ and the $+1$ 
eigenvector is $\vec y \sim (0,1/\sqrt{2},1/\sqrt{2})$. A fluctuation along 
the fundamental string $(1,0,0)$ is orthogonal to $\vec y$. It therefore has 
eigenvalue $-1$ and satisfies a Dirichlet boundary condition as
one would expect for transverse fluctuations of a fundamental string 
attached to a D-string.

It is also easy to check that the transmission and reflection amplitudes 
of three-string junctions transform appropriately under the usual $SL(2,Z)$ 
transformations of type IIb theory,
\bea
\label{sltwo}
\tau \rightarrow \tau ' &=& {a\tau +b \over c\tau +d}\,, \nonumber \\
&{ }&  \\
\left( \begin{array}{c}
p \\ q
\end{array} \right)
\rightarrow 
\left( \begin{array}{c}
p' \\ q'
\end{array} \right)
&=&
\left( \begin{array}{cc}
\phantom{-}a & -b \\ -c & \phantom{-}d 
\end{array} \right)
\left( \begin{array}{c}
p \\ q
\end{array} \right) \,.  \nonumber
\eea
where $a,b,c,d$ are integers satisfying $ad-bc=1$.
To prove this, one first shows that 
\be
p'+q'\tau' = {p+q\tau \over c\tau +d} \,,
\label{pqprime}
\ee
and then observes that the S-matrix entries in (\ref{smat}) 
only depend on a ratio of first order expressions involving the different string 
tensions.  It follows that the factors of 
$\sqrt{\vert {\rm Im}\,\tau'\vert}\, \vert c\tau+d\vert$ cancel and the form of
the S-matrix is left invariant by the $SL(2,Z)$ transformation (\ref{sltwo}).
This does not appear to be a very restrictive test of the structure 
of the scattering matrix, but it is not without content. 

\section{In-Plane Scattering at a Junction}

One can also consider fluctuations in the plane of the three-string 
junction.  This case is somewhat more complicated to deal with than
the out-of-plane fluctuations because an in-plane fluctuation that
is transverse with respect to one of the strings induces motion at the
junction that is both transverse and longitudinal with respect to the
other two strings. The longitudinal fluctuations are unphysical and 
will eventually be eliminated, but the boundary conditions are easier 
to describe if they are left in temporarily. As an aside, we note that
in the worldvolume gauge theory approach to this problem, longitudinal
displacements are accounted for by the worldvolume gauge field $A_\mu(x,t)$.
The fact that each string terminates at the junction means that the gauge
field cannot be completely eliminated by a choice of gauge. For
each string, there remains a degree of freedom corresponding to the 
longitudinal position of the end of the string and when several strings
are coupled at a junction, this degree of freedom plays an important role
in the scattering dynamics. The treatment of in-plane scattering given
below is not directly derived from the coupled worldvolume gauge theory
approach but gives, we believe, the same result a more formal treatment
would give. 

We again take the strings to lie in the complex plane with the equilibrium 
position of the junction at $z=0$ and $x_i>0$ measuring the distance 
from the junction along a given string. 
The in-plane transverse and longitudinal fluctuations on each string can 
be conveniently combined into a single complex-valued field 
$z(x,t)=\chi(x,t)+ i\varphi(x,t)$ where $\chi$ and $\varphi$ are the 
longitudinal and transverse fluctuations, respectively. 
Consider first a string extended along the positive real axis.  
Its endpoint is at 
\be
z(x=0) = \chi(0) + i \varphi(0) \,,
\ee
where $\chi$ is the longitudinal fluctuation and $\varphi$ is transverse.
To rotate to a given $(p,q)$ string one simply multiplies by the 
appropriate phase,
\be
z_i =e^{i\theta_i} ( \chi_i+i \varphi_i) .
\ee
Continuity at the junction gives two complex equations:
\be
z_1(0)=z_2(0)=z_3(0) .
\label{zcont}
\ee
This amounts to four real valued equations, of which three can be used
to eliminate the unphysical longitudinal fluctuations $\chi_i$, leaving
behind a single linear condition on the transverse fluctuations, which
reads:
\be
0=\varphi_1(0) \sin{\theta_{23}}+\varphi_2(0) \sin{\theta_{31}}+
        \varphi_3(0) \sin{\theta_{12}} ,
\label{ipcont}
\ee
where $\theta_{ij}\equiv \theta_i-\theta_j$.

The remaining two boundary conditions come from tension balance
at the junction.  When the junction is perturbed the strings will come
into it at angles that differ somewhat from the equilibrium angles, and
the condition for tension balance becomes 
\begin{equation}
\label{tensionbal}
0=\sum_{i=1}^3 e^{i(\theta_i+\delta\theta_i)} t_i .
\end{equation}
For small fluctuations the angles $\delta\theta_i$ will be small and to
leading order they only depend on the transverse fluctuations.  In fact,
they are simply given by the slope of the transverse fluctuation field
at the string endpoints:
\be
\delta\theta_i \approx \varphi_i '(0) .
\ee
Expanding the tension balance condition (\ref{tensionbal}) to first order
in fluctuations gives the linear relation
\be
0=\sum_{i=1}^3 e^{i\theta_i} t_i \, \varphi_i '(0) .
\ee
This is a complex equation and the real and imaginary parts give the
remaining two boundary conditions that we need in order to determine
the S-matrix relating the in and out parts of the transverse fluctuation 
fields $\varphi_i$. 
 
After some straightforward algebra one finds the following expression:
\be
S = {\bf 1} - {2\over D}\left(
\begin{array}{ccc}
t_2 t_3 \sin^2 \theta_{23} & 
t_2 t_3 \sin \theta_{23}\sin \theta_{31} &
t_2 t_3 \sin \theta_{12}\sin \theta_{23}  \cr
t_1 t_3 \sin \theta_{23}\sin \theta_{31} & 
t_1 t_3 \sin^2 \theta_{31} &
t_1 t_3 \sin \theta_{31}\sin \theta_{12}  \cr
t_1 t_2 \sin \theta_{12}\sin \theta_{23} &
t_1 t_2 \sin \theta_{31}\sin \theta_{12} &
t_1 t_2 \sin^2 \theta_{12} 
\end{array}
\right) ,
\label{longsmat}
\ee
where $D=t_1 t_2 \sin^2 \theta_{12} + t_2 t_3 \sin^2 \theta_{23} 
+ t_3 t_1 \sin^2 \theta_{31}$. It is easy to see that this matrix has two
$+1$ eigenvectors spanning the space orthogonal to 
$(\sin\theta_{23},\sin\theta_{31},\sin\theta_{12})$ ({\it i.e.} satisfying
(\ref{ipcont})) and one $-1$ eigenvector proportional to 
$(t_2t_3\sin\theta_{23},t_1t_3\sin\theta_{31},t_1t_2\sin\theta_{12})$.
Once again, the zero modes 
of the system must belong to the $+1$ eigenvalues. Simple trigonometry shows 
that the defining condition for these eigenvectors is satisfied by any bodily
translation of the the string junction in its plane. These are precisely the
two zero modes we would expect to find. 
%The condition is similar to, but not identical to (\ref{ipcont}), 
%which is a condition on the net displacement,
%under the sum of the incoming and outgoing waves, of the junction.

Energy is conserved as a matter of course in the scattering process for the 
same reasons as before. The same logic as for the out-of-plane fluctuations
leads us to consider rescaled fields $\hat \phi_i=\sqrt{t_i} \phi_i$ 
and the rescaled S-matrix, $\hat S =\sqrt{t}S\sqrt{t}^{-1}$.
Carrying out the rescaling on (\ref{longsmat}), we get
\be
\hat S = {\bf 1} - 2 \vec z\otimes \vec z \,,
\ee
where $\vec z$ is the unit three-vector
\be
\vec z = {1\over\sqrt{D}}
(\sqrt{t_2t_3}\sin \theta_{23} ,\sqrt{t_1t_3}\sin \theta_{31},
\sqrt{t_1t_2}\sin \theta_{12})  \,.
\ee
This is again an orthogonal matrix which squares to the identity and its
eigenvectors are orthogonal.

We can again perform a simple check by considering a fundamental string 
attached to a D-string, such that $t_1<<t_2\sim t_2$, and
$\vec z\sim (0,1/\sqrt{2},1/\sqrt{2})$.  A fluctuation on
the fundamental string $(1,0,0)$ is orthogonal to $\vec z$ and has eigenvalue 
$+1$.  It thus satisfies a Neumann boundary condition as expected.

The argument for $SL(2,Z)$ covariance of the scattering amplitudes proceeds
in much the same way as for the out-of-plane fluctuations. The only new twist
is that the S-matrix now depends on the angles between the strings as well as
their scalar tensions.  It follows from (\ref{pqprime}) that the orientation 
angle of a $(p,q)$ string transforms under $SL(2,Z)$ as 
$\theta'_i=\theta_i-\arg (c\tau+d)$. 
Since the $S$-matrix only involves differences of angles and is 
homogeneous in powers of the scalar tension, it manifestly maintains its 
form under $SL(2,Z)$ transformations. Again, the test does not seem very 
restrictive, but it is not entirely trivial that it is met.

\section{Higher Order String Junctions}

Our treatment of scattering at a three-string junction can be generalized 
to planar $n$-string junctions, with $n\geq 4$.  An $n$-string junction 
preserves $\nu=1/4$ supersymmetry when the strings that emerge from it all
lie in a single two-dimensional plane and the charge conservation and 
tension balance conditions are satisfied,
\be
\label{nconserve}
0=\sum_{i=1}^n q_i = \sum_{i=1}^n p_i = \sum_{i=1}^n e^{i\theta_i}t_i ~.
\ee
As far as we can
tell, such objects should exist in type II superstring theory.  Consider, for
example, the IR limit of a general string network formed out of a collection
of three-string junctions with $n$ external strings attached.  For wavelengths
large compared to the size of the network the physics will be that of an
$n$-string junction with appropriate matching conditions on fluctuations.

Let us first consider out-of-plane modes 
$\phi_i(t,x_i)=e^{-i\omega t} \tilde \phi_i(x_i)$, 
where $i=1,\ldots,n$ labels the strings that attach to the junction.  Continuity
at $x_i=0$ gives rise to $n-1$ equations,
\be
\label{ncont}
\tilde \phi_1(0)=\tilde \phi_2(0)=\ldots =\tilde \phi_n(0) \,,
\ee
and vertical tension balance adds the equation
\be
\sum_{i=1}^n t_i\, \tilde\phi_i '(0) \,.
\ee
The resulting S-matrix relating in- and out-modes generalizes the answer for
a three-string junction (\ref{smat}) in a straightforward way,
\be
{\bf S}=-{\bf 1} +
{2\over \sum_1^n t_i}
\left(\begin{array}{ccc}
t_1& \cdots & t_n\cr
\vdots & { }& \vdots\cr
t_1&\cdots & t_n
\end{array}\right) \,.
\label{nsmat}
\ee
The eigenvalues of this S-matrix are $(1,-1,\ldots,-1)$ for any $t_i$.
There is a single linear combination of incoming modes that satisfies a Neumann
condition at the junction and all orthogonal combinations satisfy Dirichlet 
conditions.  The Neumann mode is the one where all the strings have equal 
excitation, {\it i.e.} $B_i=b$ for all $i=1,\ldots,n$ and some constant $b$.
In the $\omega\rightarrow 0$ limit this mode describes a translational zero mode
that uniformly moves the string junction in a direction perpendicular to its 
plane. There are a total of seven such zero modes for a junction embedded in 
$9+1$ dimensional spacetime.

For in-plane scattering at an $n$-string junction we can generalize the 
discussion of the previous section in the obvious way.  Continuity at the 
junction gives $n-1$ complex equations,
\be
\label{nzets}
z_1(0)=z_2(0)=\ldots =z_n(0)  \,.
\ee
We use $n$ real valued equations to eliminate the unphysical longitudinal 
fluctuations $\chi_i$, leaving behind $n-2$ linear conditions on the transverse
fluctuations,
\be
0=\varphi_{i-1}(0) \sin{\theta_{i,i+1}} +\varphi_i(0) \sin{\theta_{i+1,i-1}}
+\varphi_{i+1}(0) \sin{\theta_{i-1,i}}  \,,
\label{nphis}
\ee
where $\theta_{i,j}\equiv \theta_i-\theta_j$.

The remaining matching conditions come from tension balance,
\be
0=\sum_{i=1}^n e^{i\theta_i} \,t_i \, \varphi_i '(0)  \,.
\ee
This complex equation can be rewritten as the following two real-valued 
equations:
\bea
\label{tenbal}
0 &=& \sum_{i=2}^n \varphi_i '(0) \, t_i \, \sin{\theta_{1,i}}  \,, \nonumber \\
0 &=& \sum_{i=1}^{n-1} \varphi_i '(0) \, t_i \, \sin{\theta_{i,n}}  \,. 
\eea

The matching conditions (\ref{nphis}) and (\ref{tenbal}) define a linear system
of $n$ equations which determines the in-plane S-matrix of the $n$-string
junction,
\be
S={\bf 1} -{2\over D} T \,.
\ee
Here $D=\sum_{i<j} t_i t_j \, \sin^2{\theta_{ij}}$, and the diagonal and 
off-diagonal elements of the $n\times n$ matrix $T$ are given by
\bea
T_{ii} &=& \sum_{\scriptstyle k<l\atop\scriptstyle k,l\neq i} 
t_k t_l \,\sin^2{\theta_{kl}}  \,, \qquad ({\rm no\ sum\ on\ }i),  \nonumber \\
T_{ij} &=& \sum_{k\neq i,j} t_j t_k \, \sin{\theta_{ik}}\sin{\theta_{kj}} \,,
\qquad (i\neq j).
\label{longsmatel}
\eea
The eigenvalues of the in-plane S-matrix are $(1,1,-1,\ldots,-1)$, which is 
once again independent of the tension of individual strings at the junction.
The $+1$, or Neumann, eigenvectors are those that are annihilated by the 
matrix $T$. Some fairly tedious trigonometry applied to (\ref{longsmatel}) 
shows that there are two independent vectors, corresponding to bodily
displacements of the junction within its own plane, that satisfy this 
Neumann condition. As before, the $\omega\to 0$ limit of the Neumann 
eigensolutions are zero modes. Taking longitudinal and transverse modes
together, we have identified a total of nine translational zero modes, just 
what we expect for a solitonic object in $9+1$ dimensions.

\section{Supersymmetry}

In the previous sections, we have computed the low-energy limit of the S-matrix
for massless bosonic excitations of a string junction. The strings support
fermionic excitations as well, and we should be able to say something about
their S-matrix. Sen \cite{sensusy} has given a supergravity argument that a 
network of $(p,q)$ strings in a plane leaves exactly eight of the thirty-two 
type-IIB supersymmetry generators unbroken and we should at least be able to 
reproduce this result. 

The low energy degrees of freedom of a $(p,q)$ string fall into eight
1+1-dimensional $(1,1)$ supermultiplets (corresponding to the eight directions
transverse to the string). In a string network, we have multiple string
segments lying in a plane and coupled to each other through boundary
conditions at their junctions. The supermultiplets corresponding to 
the seven displacements perpendicular to the plane are decoupled from each other
and completely equivalent. The single supermultiplet corresponding to 
displacements in the plane of the network is decoupled from, and inequivalent
to, the rest. The question is whether, in the presence of the junctions,
each of these eight theories manages to have one surviving supersymmetry. 
If they do, that would give the expected total of eight supersymmetries. 
 
Let us consider fluctuations in a particular one of the directions
transverse to the plane of the network. We know that the low-energy dynamics 
of an individual $(p,q)$ string is described by a two-dimensional massless 
Majorana fermion 
$\psi$ plus a real scalar $\phi$. Because the fields are massless, they can be 
decomposed into left- and right-moving supermultiplets 
$(\phi_+,\psi_+)$ and $(\phi_-,\psi_-)$. The $\psi_\pm$ are now 
single-component anticommuting objects and the $(1,1)$ supersymmetries 
basically exchange the $\psi_\pm$ with the $\phi_\pm$. A junction imposes a 
boundary condition which couples together the left- and right-movers on 
different strings. The question is whether this can be done in a 
supersymmetric fashion. 

The answer is pretty trivially yes. The first step is to rewrite the
boundary conditions (\ref{match},\ref{tbal}) on transverse bosonic coordinates 
in terms of the left- and right-moving components defined by
$\phi(\tau,\sigma)=\phi_+(\tau+\sigma)+\phi_-(\tau-\sigma)$. The boundary
condition at $\sigma=0$ can be recast as a relation between fields of 
the two different chiralities:
\be
\phi_1^+-\phi_2^+=-\phi_1^-+\phi_2^-,\quad 
	\phi_2^+-\phi_3^+=-\phi_2^-+\phi_3^-,\qquad
\sum_{i=1}^3 t_i{\phi_i^+}^\prime = \sum_{i=1}^3 t_i{\phi_i^-}^\prime~.
\label{chirbc}
\ee
Modulo a possible zero-mode subtlety, we can integrate the last of these 
to eliminate the derivative and recast the boundary condition as a simple
linear map between the $\phi_i^+$ and $\phi_i^-$. By passing to the hatted 
fields $\hat\phi_i=\sqrt{t_i}\phi_i$, we can then diagonalize the boundary
conditions by an orthogonal transformation:
\bea
\sum_i l_i\hat\phi_i^+& =& -\sum_i l_i\hat\phi_i^- ~,~~~
	\vec l = {(\sqrt{t_2},-\sqrt{t_1},0)\over \sqrt{t_1+t_2}} ~,
\nonumber\\
\sum_i m_i\hat\phi_i^+&=&-\sum_i m_i\hat\phi_i^- ~,~~~
	\vec m = {(\sqrt{t_3t_1},\sqrt{t_3t_2},-(t_1+t_2))\over
        \sqrt{t_1+t_2}\,\sqrt{t_1+t_2+t_3}} ~,
\nonumber\\
\sum_i n_i\hat\phi_i^+&=&+\sum_i n_i\hat\phi_i^- ~,~~~
	\vec n = {(\sqrt{t_1},\sqrt{t_2},\sqrt{t_3})\over
	\sqrt{t_1+t_2+t_3}} ~,
\nonumber\\
\vec l^2=\vec m^2&=&\vec n^2=1, \qquad 
	\vec l\cdot\vec m=\vec l\cdot\vec n=\vec n\cdot\vec m=0 ~.
\label{diagbc}
\eea 
The content of this is that the fields $\vec l\cdot\hat\phi$ and 
$\vec m\cdot\hat\phi$ satisfy Dirichlet boundary conditions while the
field $\vec n\cdot\hat\phi$ satisfies a Neumann boundary condition.
Now we can impose equivalent boundary conditions on the fermions by replacing 
the $\hat\phi^\pm$ in these equations by the single-component anticommuting
$\hat\psi^\pm$. Since the bulk supersymmetry transformation is implemented
by swapping $\phi^\pm$ with $\psi^\pm$, this is the only possible supersymmetric
boundary condition. Indeed, the allowed boundary conditions
for Majorana fermions are just $\psi^+=\pm\psi^-$ and we are choosing to
impose precisely such boundary conditions on orthogonal linear combinations
of $\hat\psi_i$. 
 
Now let's think about supersymmetry of a network of junctions. In a network, 
for each leg, $i$, we have one $N=1$ supercharge: 
$$
Q_i=\int_0^{l_i}d\sigma(\hat\psi^+_i\hat{\phi^+_i}^\prime +
	\hat\psi^-_i\hat{\phi^-_i}^\prime) ~. 
$$
Possible non-conservation of $Q_i$ comes from boundary terms:
$$
{d\over dt} Q_i = (\hat\psi^+_i\hat{\phi^+_i}^\prime-
	\hat\psi^-_i\hat{\phi^-_i}^\prime)|^{l_i}_0~.
$$
This would vanish if Neumann/Dirichlet boundary conditions
$\hat\phi_i^+=\pm\hat\phi_i^-$ and $\hat\psi_i^-=\pm\hat\psi_i^-$
were applied directly to the fields on a given leg. That is not the situation
in a network, however. In the network supercharge $Q=\sum_iQ_i$, the
boundary terms organize themselves into a sum of contributions from the 
various junctions:
$$
{d\over dt} Q_{junc} = 
	\sum_{i=1}^3  (\hat\psi^+_i\hat{\phi^+_i}^\prime-
       		 \hat\psi^-_i\hat{\phi^-_i}^\prime)
$$
where the sum is over the legs that meet at the junction in question
(and the fields are evaluated at the junction).
By orthogonality, we can rewrite this in terms of the fields which diagonalize
the boundary conditions:
\bea
{d\over dt} Q_{junc}& = &
       (\vec l\cdot \hat\psi^+_i{\vec l\cdot\hat \phi^+_i}^\prime-
        \vec l\cdot \hat\psi^-_i{\vec l\cdot\hat\phi^-_i}^\prime) 
+ (\vec m\cdot \hat\psi^+_i{\vec m\cdot\hat \phi^+_i}^\prime-
        \vec m\cdot \hat\psi^-_i{\vec m\cdot\hat\phi^-_i}^\prime)
\nonumber \\ 
& & 
+ (\vec n\cdot \hat\psi^+_i{\vec n\cdot\hat \phi^+_i}^\prime-
        \vec n\cdot \hat\psi^-_i{\vec n\cdot\hat\phi^-_i}^\prime)~.
\eea
This is a sum of terms which individually vanish because the relevant fields
satisfy supersymmetric Neumann or Dirichlet boundary conditions. Therefore 
the sum automatically vanishes. This argument works at any junction in the 
network, even if the tensions involved and the orthogonal projections are 
different at different junctions. Consequently, for each 
transverse direction, we have exactly one supercharge. Much the 
same argument goes through for the in-plane excitations as well and we conclude
that the network has a total of eight supercharges, as expected.
 
\section{Excitations on String Lattices}

In this section we will discuss the dynamics of a web of junctions that 
forms a periodic lattice. 
We will consider the system obtained by placing two identical triple 
junctions on a 2-torus and connecting up strings of like charge. 
The setup, then, is three different strings, of three different lengths 
(more or less freely adjustable by adjusting the modulus of the torus) 
coupled together at two mirror image triple junctions, as shown in Figure~2. 
The static geometry of this setup is determined by the string tensions and 
windings. We would like to determine the low-lying excitations of this 
system in order to assess how its entropy, conformal invariance and 
supersymmetry properties would differ from those of the more familiar 
simply-wound D-string. 

\vskip .5cm
\vbox{
{\centerline{\epsfxsize=2.4in \epsfbox{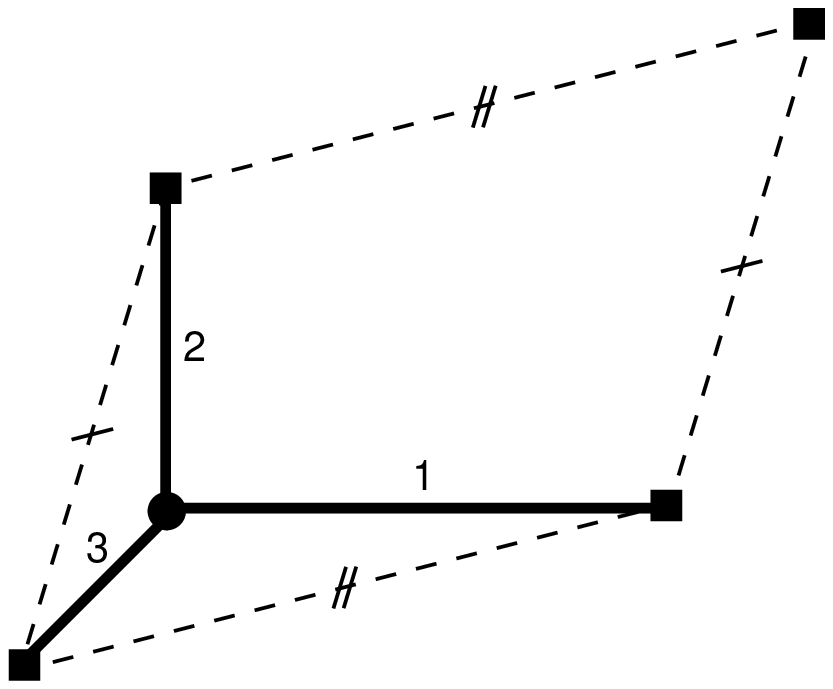}}}
{\centerline{FIGURE 2: A string lattice with two junctions per 
unit cell.}} 
{\centerline{Opposite sides of the parallelogram are identified 
so that one}}
{\centerline{of the junctions is represented four times in the 
diagram.}}
}
\vskip .5cm

Our first exercise will be to construct the eigenvalue condition for 
out-of-plane bosonic excitations. The three string segments joining the 
two junctions have scalar tensions and lengths $t_i, l_i$ $(i=1,2,3)$. 
By varying the torus parameters, the lengths can be made pretty much 
arbitrary, so we will keep them general for now. We can regard the 
displacement field on each of the three strings as an independent real 
free massless field $\phi_i(x_i,t)$ living on its own line segment 
$0<x_i<l_i$. The most general disturbance of frequency $\omega$ has been
written down in (\ref{modes}), but, in order to study the effect of the
second junction (at $x_i=l_i$), it is helpful to also write the same fields
in terms of ``conjugate'' variables $y_i=l_i-x_i$:
\bea
\phi_i(x_i,t) &=& Re\{(A_i e^{i\omega x_i} + 
	B_i e^{-i\omega x_i}) e^{-i\omega t}\}\cr 
	&=& 
Re\{(B_ie^{-i\omega l_i} e^{i\omega y_i} + 
	A_i e^{i\omega l_i} e^{-i\omega y_i}) e^{-i\omega t}\} ~.
\label{torfield}
\eea
The parameters $A_i$, $B_i$ and $\omega$ are constrained by the boundary 
conditions imposed by the S-matrix (\ref{smat}) at the two junctions:
\bea
\label{latmat}
\vec A =& S\cdot\vec B\qquad\qquad &{\rm at}~x_i=0 ~, \nonumber \\
\qquad P\cdot\vec B=& S\cdot P^*\cdot\vec A\qquad &{\rm at}~y_i=0 ~, 
\eea
where $P={\rm diag}(e^{-i\omega l_1},e^{-i\omega l_2},e^{-i\omega l_3})$ is 
a phase matrix which accounts for the different propagation phases along
the different length legs between the two vertices. 

It turns out that these equations are separately satisfied by the real and
imaginary parts of the mode amplitudes.  This is a consequence of the fact
that $S$ is a real matrix and it allows us to focus on, say, real parts of
$A_i$ and $B_i$ only.  Furthermore, since the problem is linear, the 
overall scale of the fields must drop out leaving only five field parameters
to determine, plus one energy, for a total of six.
With some uninspiring algebra, the system 
can be boiled down to a single eigenvalue condition on the energy:
\be
0 = (\sum_{i=1}^3 t_i^2)s_1s_2s_3 + 2 t_1t_2s_3(1-c_1c_2) 
	+2 t_1t_3s_2(1-c_1c_3) + 2 t_2t_3s_1(1-c_2c_3) \,,
\label{spect}
\ee
where $s_i=\sin{\omega l_i}, c_i=\cos{\omega l_i}$.  
For any given eigenvalue $\omega$ that satisfies this condition
there is a mode vector $\vec A$ which is, in general, a linear superposition
of the three eigenvectors of the S-matrix at one of the vertices. This 
means, roughly speaking, that the general mode satisfies neither D nor
N boundary conditions. This has implications for conformal invariance,
as we will discuss. 

It is hard to say anything general about the transcendental condition
(\ref{spect}) other than that the number and spacing of eigenvalues is 
roughly what is expected.  We can get some insight by studying some special 
cases. First, let the three string lengths be equal, in which case the 
eigenvalue condition reduces to $\sin{\omega l}=0$, which gives the energy 
levels of an NN or DD open string. This is just right, because the whole 
problem can be trivially diagonalized, by taking orthogonal linear 
combinations of the $\phi_i$, into one NN and two DD open strings of 
length $l$.  It is also worth noting that we always have 
one zero mode $\omega=0$, no matter what the lengths and tensions are. 
Another special case of interest is when we have one fundamental string 
and two very heavy D-strings.  Then the tensions are related by 
$t_1<< t_2\sim t_3$, and in this limit the eigenvalue condition becomes
\be
\sin{\omega l_1}\left(1-\cos{\omega(l_2+l_3)}\right)=0 \,.
\ee
This gives the spectrum of the DD open string on $l_1$ plus that of a closed 
string (left-movers plus right-movers) on $l_2+l_3$. This is precisely what 
you would expect from the Polchinski boundary condition approach to the 
dynamics of fundamental strings attached to D-strings. 

A similar exercise gives us the equation for the spectrum of in-plane 
disturbances,
\begin{eqnarray}
\label{ipspec}
0 &=& ( (\tau_{12})^2 + (\tau_{23})^2 + (\tau_{31})^2) s_1 s_2 s_3 
+ 2 \tau_{12} \tau_{31} s_1 (1-c_2 c_3) \nonumber \\
&  &\qquad + 2 \tau_{12} \tau_{23} s_2 (1-c_1 c_3)
+ 2 \tau_{23} \tau_{31} s_3 (1-c_1 c_2)\,,
\end{eqnarray}
where  $\tau_{ij} = t_i t_j sin^2(\theta_{ij})$ and $\theta_{ij}$ is the
angle between the the $ij$ string  pair.  This is similar, but by no means
identical, in structure to the out-of-plane spectral equation. By taking
appropriate limits, we see that it does the right thing for the following 
special cases: (1) all lengths equal, and (2) the weak coupling limit of a 
fundamental string attached to a D-string. 

Let us discuss the issue of conformal invariance. The individual junctions
impose conformal boundary conditions (D or N) on orthogonal linear
combinations of fields.  When two junctions are connected up in the most 
general fashion, however, one does not see towers of equally-spaced levels 
corresponding to the Verma modules of the two-dimensional conformal group. 
Only in very special cases, such as those discussed above, do we find 
standard conformal towers. At some level this is to be expected: strict 
conformal invariance is after all a property of tree level string theory while 
the whole point of the string junction construction is to include physics that 
is non-perturbative in $g_s$ (through the string tensions at least). 
On the other hand, we might hope to use this setup as a toy model for 
understanding how the dynamical principle of conformal invariance gets 
generalized beyond string tree level, but we have not seen how to exploit
this possibility concretely. 

Our formalism can also be used to study the band structure of the string
lattice.  When we determined the lattice spectrum above, we imposed 
periodic conditions on the fields across a unit cell of the lattice.
An infinite lattice also supports modes with wavelengths that span 
several unit cells.  A mode which comes back to itself after $n$ lattice
spacings only has to be periodic up to a phase across a single lattice
spacing.  This phase can be any $n$-th root of unity, 
$e^{2\pi im/n}$, and since $n$ can take arbitrarily large values on an
infinite lattice the phase is in fact completely unrestricted.

It is straightforward to allow for this phase freedom in our equations.
With the lattice strings labelled as in Figure~2, this is achieved by
a suitable modification of the matching conditions at one of the two
string junctions:
\bea
\label{platmat}
\vec A =& S\cdot\vec B\qquad\qquad &{\rm at}~x_i=0 ~, \nonumber \\
\qquad P\cdot\alpha\cdot\vec B
=& S\cdot P^*\cdot\alpha\cdot\vec A\quad &{\rm at}~y_i=0 ~, 
\eea
where the matrix $\alpha ={\rm diag}(e^{i\alpha_1},e^{i\alpha_2},1)$, 
with $0\leq \alpha_1,\alpha_2 \leq 2\pi$, contains the phase information
along the two independent lattice vectors.  When general phases are
inserted, the real and imaginary parts of the mode amplitudes no 
longer decouple in the matching conditions but at the end of the day
the system of equations still reduces to a single real equation
for the mode energy.  For out-of-plane fluctuations the spectral 
equation (\ref{spect}) generalizes to
\bea
\label{pspec}
0 &=& (\sum_{i=1}^3 t_i^2)s_1s_2s_3 
+ 2 t_1t_2s_3(\cos{(\alpha_1-\alpha_2)}-c_1c_2) \nonumber \\
& &+2 t_1t_3s_2(\cos{\alpha_1}-c_1c_3) 
+2 t_2t_3s_1(\cos{\alpha_2}-c_2c_3) ~.
\eea
At generic parameter values it is hard to extract useful information
from this equation, but the qualitative behavior is that
allowed values of $\omega$ fall into continuous bands as the phase
angles $\alpha_1$ and $\alpha_2$ range from $0$ to $2\pi$.  Let us 
illustrate this for the special case when the three strings in the
unit cell all have the same length, $l_1=l_2=l_3\equiv l$, so that
$c_1=c_2=c_3=\cos{\omega l}$ and $s_1=s_2=s_3=\sin{\omega l}$.  
With this simplification the spectral equation (\ref{pspec}) can be
solved explicitly:
\be
\label{bands}
\sin{\omega l}=\pm {2\sqrt{t_1t_2\sin^2({\alpha_1-\alpha_2\over 2})
                + t_1t_3\sin^2({\alpha_1\over 2})
                + t_2t_3\sin^2({\alpha_2\over 2})} 
                \over t_1+t_2+t_3} ~.
\ee
The absolute value of the right hand side is always less than or equal
to one so that the allowed values of $\omega$ lie in continuous bands
centered around $\omega =n\,\pi$  for $n\in {\bf Z}$.  The bands will
overlap with each other if and only if the right hand side of 
(\ref{bands}) equals $\pm1$ for some value of $\alpha_1$ and $\alpha_2$.
This can only happen if the scalar tension $t_i$ of one of the strings
equals the sum of the other two.  In this case the strings are all 
parallel and the string lattice is degenerate.  On a non-degenerate
lattice the out-of-plane excitations will always have finite band gaps.
If one of the strings is a fundamental string the band gap will be 
finite but very narrow at weak string coupling.

For generic values of the string lengths the detailed analysis of the
band spectrum becomes more complicated but the qualitative features
are unchanged.  The same method can be used to derive the 
appropriate generalization of the in-plane spectral equation 
(\ref{ipspec}) and obtain from it the band spectrum of in-plane modes.

Rey and Yee \cite{soojong} have also studied aspects of propagation
on string lattices. They claim to find an `evanescent bound state' 
for which we see no evidence. It may be that their state is an approximate 
manifestation of the opening of the band gap which we find to be a generic 
lattice feature, but we have not tracked down the precise correspondence. 

\section{Discussion}

We have explored the infrared dynamics of string junctions and string
networks.  Only the most basic features of the relativistic
string entered into our considerations, so our results are presumably
quite reliable at low energy. If anything, it 
is surprising how much structure the system has, given how little goes 
into it in the way of dynamical information.  

One important issue is the energy scale beyond which the simple 
viewpoint adopted in the present paper becomes inadequate.  Clearly,
when mode energies approach the string scale we expect particle 
production in the worldsheet theory at a string junction, in which
case the S-matrix can no longer be determined by its action on 
one-particle states alone.  At weak string coupling, $g_s<<1$, there
is a lower energy scale where new physics enters.  The worldsheet
theory of strings carrying $q$ units of R-R charge is an $N=8$ 
supersymmetric $U(q)$ gauge theory and the $(p,q)$ string corresponds
to a vacuum in this theory with $p$ quarks in the fundamental 
representation of $U(q)$ placed at infinity \cite{witten}.  
In this vacuum, massless excitations carry $U(1)$ charge but are
$SU(q)$ singlets, as is required by the $SL(2,Z)$ duality of type IIB
string theory.  The mass gap for the non-abelian degrees of freedom 
is $\Delta m\sim g_s/\sqrt{\alpha'}$, and when our mode energies 
approach this scale the worldsheet dynamics will become non-trivial.
For an isolated string junction, our results will be valid for modes
with wavelengths longer than $\sqrt{\alpha'}/g_s$, and similarly, our
spectral equations for string lattices will hold provided the lattice 
spacing is sufficiently large, $l>\sqrt{\alpha'}/g_s$.  It would be 
interesting to identify the leading effects of the non-abelian worldsheet 
structure as the relevant energy scale is approached from below.

Our results lay the groundwork for attacking at least two more questions 
of potential interest. The first has to do with
the nature of the conformal invariance that survives in a stringy system
beyond the string tree level. The second has to do with the properties
of black hole analog states constructed by wrapping brane networks, rather
than the usual D-branes, about compact tori. Do we find a new class of
extremal black holes and their near-extremal relatives, or are we seeing
old friends in new clothes? We hope to examine these questions in future
investigations.

\section*{Acknowledgements}

This work was supported in part by the US Department of Energy 
grant DE-FG02-91ER40671 and by National Science Foundation grant
PHY96-00258.

\end{document}